 \documentclass[final,5p,times,twocolumn,authoryear]{elsarticle}

\usepackage{booktabs}
\usepackage{verbatim}

\usepackage{amssymb}
\usepackage{amsthm}
\usepackage{amsmath}

\usepackage{longtable}
\usepackage[CaptionAfterwards]{fltpage}

\journal{Earth and Planetary Science Letters}

\begin{document}

\begin{frontmatter}



\title{Formation, stratification and mixing of the cores of Earth and Venus}


\author[label1,label2]{Seth A. Jacobson}
\author[label1]{David C. Rubie}
\author[label3]{John Hernlund}
\author[label2]{Alessandro Morbidelli}
\author[label4]{Miki Nakajima}
\address[label1]{Universit{\"a}t Bayreuth, Bayerisches Geoinstitut, D-95440 Bayreuth, Germany}
\address[label2]{Observatoire de la C{\^o}te d'Azur, Laboratoire Lagrange, Bd. de l'Observatoire, CS 34229, F-06304 Nice Cedex 4, France}
\address[label3]{Tokyo Institute of Technology, Earth-Life Science Institute, 2-12-1-IE-1 Ookoyama, Meguro-ku, Tokyo, 152-8550, Japan}
\address[label4]{Carnegie Institution of Washington, Department of Terrestrial Magnetism, 5241 Broad Branch Rd., NW, Washington, DC, 20015-1305, USA}

\begin{abstract}
Earth possesses a persistent, internally-generated magnetic field, whereas no trace of a dynamo has been detected on Venus, at present or in the past, although a high surface temperature and recent resurfacing events may have removed paleomagnetic evidence.
Whether or not a terrestrial body can sustain an internally generated magnetic field by convection inside its metallic fluid core is determined in part by its initial thermodynamic state and its compositional structure, both of which are in turn set by the processes of accretion and differentiation.
Here we show that the cores of Earth- and Venus-like planets should grow with stable compositional stratification unless disturbed by late energetic impacts.
They do so because higher abundances of light elements are incorporated into the liquid metal that sinks to form the core as the temperatures and pressures of metal-silicate equilibration increase during accretion.
We model this process and determine that this establishes a stable stratification that resists convection and inhibits the onset of a geodynamo.
However, if a late energetic impact occurs, it could mechanically stir the core creating a single homogenous region within which a long-lasting geodynamo would operate.
While Earth's accretion has been punctuated by a late giant impact with likely enough energy to mix the core (e.g. the impact that formed the Moon), we hypothesize that the accretion of Venus is characterized by the absence of such energetic giant impacts and the preservation of its primordial stratifications.
\end{abstract}

\begin{keyword}
planet formation \sep planetary differentiation \sep core formation \sep early Earth \sep early Venus \sep geodynamo
\end{keyword}

\end{frontmatter}


\section{Introduction}

Earth's magnetic field is generated inside its convecting fluid outer core, and paleomagnetic evidence indicates that it has persisted since at least 4.2~Ga \citep{Tarduno:2015fy}.
Seismological probing of the core suggests that it consists mostly of iron and nickel with approximately 10 wt\% light elements (i.e., an uncertain mixture of Si, O and S and potentially others such as H and C) \citep[see][for review]{Poirier:1994jh}.
Besides the possible stratified layers in the uppermost \citep{Helffrich:2010cy,Buffett:2014jd} and lowermost outer \citep{Gubbins:2008bu} core, the average structure is consistent with isentropic compression of a homogeneous liquid \citep{Hirose:2013io}.
Dynamical constraints suggest that the bulk of Earth's outer core is exceptionally well-mixed, exhibiting density fluctuations of order one part in a billion or less relative to an hydrostatic equilibrium profile \citep{Mandea:2012hy}.
However, it is not known how Earth's core achieved this high degree of homogeneity and whether such a high degree of homogeneity is expected for all terrestrial planets.

Terrestrial planets like Earth grow from a series of accretion events characterized by collisions with bodies, most of which had cores of their own.
In other words, Earth's core is not created in a single stage but from a series of core forming events \citep[multistage core formation is reviewed in][]{Rubie:2016kx}.
A core formed over multiple stages is not in chemical equilibrium with the mantle since each core addition equilibrates with only part of the mantle \citep{Deguen:2011ig,Rubie:2015fj}.
Moreover, the core is not necessarily chemically homogenous or isentropic at the end of planet formation.
Only further processing within the core removes the signatures of multistage core formation and creates the practically homogenous core observed today.

In order to determine the chemical state of the core during and after planet formation, we linked a terrestrial planet formation model, a planetary differentiation model, and a core growth model together (Section~2).
From these linked models, we obtained thermal and compositional profiles of the cores of Earth and Venus.
We find that the memory of multistage core formation remains as a distinct compositional stratigraphy within the core.
While convection may occur within certain layers, some boundaries between layers resist convection, require conductive heat transport, and create multiple convective cells within the core.
However, we also determined that the density profile of the core has a strong dependence on the efficiency of impact driven core mixing (Section~3).
If the impact energy from planetary accretion events is efficiently converted into turbulent mixing of the core, then the core is mechanically mixed and homogenized.
Otherwise, the density structure is preserved within the core.
As a consequence, a planet with this preserved stable stratification may not be able to produce an Earth-like geodynamo (Section~4).
We hypothesize that such an internal structure is still present in Venus, whereas the core of Earth was sufficiently mixed by the Moon-forming impact (Section~5).

\section{Establishing the structure of the core from accretion}
In order to understand the growth of Earth's core, we used previously published simulations of the growth of Earth from the accumulation of planetesimals and planetary embryos out of the terrestrial protoplanetary disk \citep{Jacobson:2014it}.
These simulations are described in detail in the supplementary information.
For clarity, we focus on the results of a well-studied simulation, 4:1-0.5-8, which is the same as that examined in \citet{Rubie:2015fj,Rubie:2016hl}.
We pass the accretion histories of each planet to a planetary differentiation model, in which we calculated the chemical evolution of each planet's mantle and core as described in \citet{Rubie:2011cr,Rubie:2015fj,Rubie:2016hl}.
This model uses data from high pressure laboratory experiments and a mass balance and element partitioning approach to calculate the composition of core forming liquids after each accretion event.
Any equilibrated metal liquid continues sinking to the core due to the high density contrast between metal and silicate, while equilibrated silicate material is mixed with the rest of the mantle.

We calculated reference core density, mass, gravity and pressure profiles using an iterative process.
After every core addition, we constructed a two-layer planet model using a pair of Murnaghan equations of state for a silicate mantle and a metallic core.
This reference density profile as a function of pressure $P$ was fitted to the mantle and the liquid outer core of the preliminary reference Earth model \citep[PREM;][]{Dziewonski:1981bz}:

\begin{equation}
\rho_\text{ref}(r) =
\begin{cases}
1669 \left( 18.89+ 5.517 P(r) \right)^{1/5.517} & \text{if $r > R_\text{CMB}$} \\
1438 \left( 195.7 + 3.358 P(r) \right)^{1/3.358} & \text{if $r \leq R_\text{CMB}$}\\
\end{cases}
\label{eqn:coredensity} 
\end{equation}

\noindent where the reference density $\rho_\text{ref}$ is measured in kg~m$^{-3}$ and the pressure $P$ is measured in GPa.
Both the mass of the planet $M$ and the mass of the core $M_\text{C}$ are known from the planetary accretion model, so from the following equation, we determined the radius of the core $R_{CMB}$ and the radius of the planet $R$.

\begin{align}
M & =  4 \pi \int_{R_\text{CMB}}^R \rho_\text{ref}(r') r'^2 dr' + M_\text{C} \\
M_\text{C} & =  4 \pi \int_0^{R_\text{CMB}} \rho_\text{ref}(r') r'^2 dr'
\end{align}

\noindent Then we used the following equations to determine the gravitational acceleration and pressure profiles.

\begin{align}
  g(r) & = \frac{4 \pi G}{r^2} \int_{0}^{r} \rho_\text{ref}(r') r'^2\ dr' \\
  P(r) & = \int_{r}^R \rho_\text{ref}(r') g(r')\ dr'
\end{align}

\noindent This iterative procedure needed an initial guess, so we used an uncompressed ($P=0$~GPa) density profile to initially calculate the core and surface radii given the core and planet mass.
We iterated through the equations above until the relative difference between consecutive density, gravity and pressure profiles added in quadrature is less than $10^{-6}$, which typically took about 10 iterations.
The core growth model calculates perturbations to this reference model due to the varying thermal and compositional properties of each core addition.

\begin{figure}
\centering
\includegraphics[width=9cm]{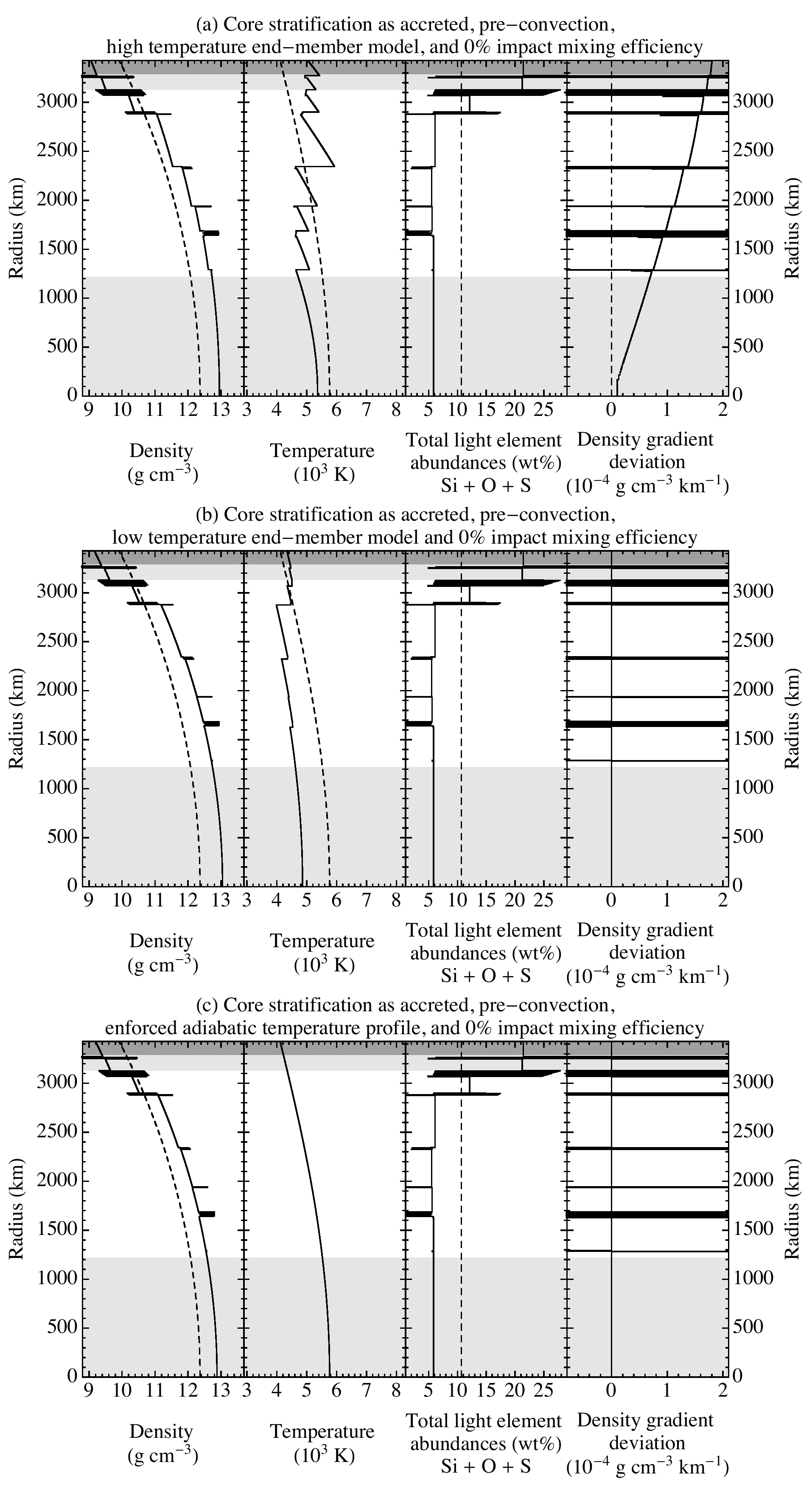}
\caption{
As solid lines, the panels show the radial profiles of the density ($\rho$; left-most panel), temperature ($T$; left-center panel), wt\% abundance of total light elements (Si, O and S; right-center panel), and the deviation of the density gradient from an isentrope ($\partial \rho /\partial r - \left. \partial \rho / \partial r \right|_S$; right-most panel) of the core of the Earth-like planet.
A completely-mixed adiabatic reference model built from a fit to the outer core of the preliminary reference Earth model \citep[PREM;][]{Dziewonski:1981bz} extrapolated to Earth's center and possessing a homogenous composition identical to that of the bulk Earth-like planet's core is shown as a dashed line.
The extent of the modern inner core is marked by a lower gray region.
The two estimates of the stratified layer thickness, an approximately 300 km stratified layer estimated from seismic modeling \citep{Helffrich:2010cy} and an approximately 140 km stratified layer estimated from geomagnetic modeling \citep{Buffett:2014jd}, are shown in gray and dark gray regions at the top of the core, respectively.
Each subfigure in this figure and in Figs.~\ref{fig:afterdensity}, \ref{fig:afterconvection}, and \ref{fig:afterimpacts} shows the same Earth-like planet with changing modeling assumptions.
All of the subfigures in this figure show the core as accreted, i.e. no evolution during or after accretion.
These three subfigures show the consequences of different assumptions regarding thermal transport during the descent of core additions and thermal transport in the core itself by examining three different end-member models as described in the text.
}
\label{fig:asaccreted}
\end{figure}

\subsection{Establishing the thermal structure of the core}
As new core forming liquids sink through the mantle, they are heated by adiabatic compression and released gravitational potential energy.
Immediately after equilibration, the metallic liquids have a temperature $T_\text{eq}$, which is approximately halfway between the peridotite solidus and liquidus at the metal-silicate equilibration pressure $P_\text{eq}$.
As this material sinks to the core-mantle boundary, it is adiabatically compressed and so heats up to a temperature at the core-mantle boundary $T_\text{CMB}$ of:

\begin{equation}
T_\text{CMB} = T_\text{eq} + \left.\frac{dT}{dP}\right|_\text{S} \left( P_\text{CMB} - P_\text{eq} \right)
\label{eqn:cold}
\end{equation}

\noindent where $P_\text{CMB}$ is the pressure at the core-mantle boundary and $\left.dT/dP\right|_\text{S} = 7.7$~K~GPa$^{-1}$ is the adiabatic temperature gradient for core fluids.
Furthermore, gravitational potential energy is released as the denser core fluids sink through the less dense silicate mantle.
If this heat is fully retained, then the temperature of the core addition when it reaches the core-mantle boundary is:

\begin{equation}
T_\text{CMB} = T_\text{eq} + \left.\frac{dT}{dP}\right|_\text{S} \left( P_\text{CMB} - P_\text{eq} \right) + \frac{g_\text{eq} r_\text{eq} - g_\text{CMB} R_\text{CMB}}{4 \pi c_P}
\label{eqn:hot}
\end{equation}

\noindent where $c_p = 825$~J~kg$^{-1}$~K$^{-1}$ is the estimated specific heat capacity at constant pressure for core fluids, $g_\text{eq}$ and $g_\text{CMB}$ are the gravitational accelerations at the radius of equilibration $r_\text{eq}$ and the core-mantle boundary $R_\text{CMB}$, respectively.
As the core continues to grow, layers already within the core continue to adiabatically compress and increase in temperature:

\begin{equation}
T = T_\text{CMB} + \left.\frac{dT}{dP}\right|_\text{S} \left( P - P_\text{CMB} \right)
\end{equation}

\noindent where $T$ is the temperature of the layer in the core at pressure $P$.

It is unclear how much of the released gravitational potential energy is retained within the sinking core addition as heat, so we examine this process in light of two end-member scenarios.
In the high temperature end-member model corresponding to Eq.~\ref{eqn:hot}, all generated heat from adiabatic compression and sinking in the gravitational potential is retained within the newly formed layer of liquid metal.
Alternatively, in the low temperature end-member model corresponding to Eq.~\ref{eqn:cold}, the new core addition is heated only by adiabatic compression; all of the released gravitational potential energy is assumed to be transported away in the silicate mantle.
Reality likely lies between the low and high temperature end-member models, however both establish a nearly isothermal core structure (see Fig.~\ref{fig:asaccreted} (a) and (b)).

These two end-members would leave the mantle, particularly at the core-mantle boundary, in different thermal states.
In the cold end-member model, the mantle would be very hot and thermal energy is unlikely to be vigorously transported across the core-mantle boundary, whereas for the hot end-member model, the mantle would be cooler and so thermal energy would more easily be conducted across this boundary.
While we introduce these figures later, we note now that in Figs.~\ref{fig:afterdensity}, \ref{fig:afterconvection}, \ref{fig:afterimpacts} and \ref{fig:venus} we use the high temperature end-member model since some released gravitational potential energy is likely transported into the core making parts of the core initially superadiabatic qualitatively similar to this model.

The thermal perturbations to the density of a core layer are calculated as:

\begin{equation}
\rho = \rho_\text{ref} \left( 1 - \alpha \left( T - T_\text{ref} \right) \right)
\end{equation}

\noindent where $\alpha = 1.5 \times 10^{-5}$ K$^{-1}$ is the coefficient of thermal expansivity for liquid iron \citep{Gubbins:2003vs} and $T_{ref}$ is the reference temperature from an isentropic model:

\begin{equation}
T_\text{ref} = T_\text{CMB,ref} + \left.\frac{dT}{dP}\right|_{S} \left( P - P_\text{CMB,ref} \right)
\label{eqn:adiabaticmodel}
\end{equation}

\noindent where $T_\text{CMB,ref} = 4100$~K and $P_\text{CMB,ref} = 135$~GPa are the reference core-mantle boundary temperature and pressure. 
Ultimately, since our conclusions rest only on differences in density between layers, they are insensitive to the chosen values of $T_\text{CMB,ref}$, $P_\text{CMB,ref}$, and $dT/ dP |_\text{S}$.

Over the age of the solar system, thermal perturbations will be conducted away, and so the sharp jumps observed in the temperature profiles (e.g. Fig. 1) would soften significantly.
However, the density changes due to variations in composition are much larger than those due to variations in temperature, so even if the temperature variation were completely disregarded and an adiabatic temperature profile was assumed to be instantaneously established, the density structure would persist.
We demonstrate this using another end-member model (see Fig.~\ref{fig:asaccreted} (c)), where we  force the growing core to always maintain an adiabatic temperature profile given by Eq.~\ref{eqn:adiabaticmodel} regardless of the new core addition's equilibrating temperature and heating history.
Comparing the density profile from all three thermal models, we see that thermal anomalies produce an insignificant effect on the density profile compared to the varying compositional structure of the core.

\subsection{Establishing the composition of the core}
The composition of the core evolves as the planet grows according to a well developed planetary accretion and differentiation model \citep{Rubie:2011cr,Rubie:2015fj,Rubie:2016hl}.
The initial composition of each planetesimal and planetary embryo contains non-volatile elements in near solar system (i.e. CI chondrite) relative abundances, while volatiles such as oxygen, sulfur and water are present in variable abundances according to radial gradients in the disk.
All embryos and planetesimals are assumed to have undergone core-mantle differentiation at the start of the simulation if they are reduced enough to contain metal, i.e. from the inner solar system.
After each accretion event, metal-silicate equilibration occurs between the dispersed metal droplets from the projectile's core and a fraction of the target's mantle \citep{Rubie:2003hq}; this fraction is determined from laboratory experiments \citep{Deguen:2011ig}.
By tracking the major element composition as well as many minor and trace elements, we modeled metal-silicate equilibration during each accretion event using a mass balance and element partitioning approach \citep[described in detail in][]{Rubie:2011cr}, which is based on laboratory determinations of metal-silicate partitioning to take into account the effects of changing composition (i.e. oxygen fugacity), pressure and temperature \citep{Mann:2009jj,Frost:2010gy,Boujibar:2014bd}.
We also modeled the process of iron sulfide segregation \citep[the model is identical to that in][]{Rubie:2016hl}, which occurs during mantle magma ocean solidification and is necessary to explain the measured low abundances of the highly siderophile elements in Earth's mantle as well as the final S content of core, but the additions themselves contain very little mass and do not substantially change the bulk composition of the core.

As the mass of the planetary embryos increase, the metal-silicate equilibration temperature and pressure increase. 
The metal-silicate equilibration temperature is taken to be approximately at the midway point between the peridotite solidus and liquidus at the equilibration pressure \citep{Rubie:2015fj}.
In order to estimate the equilibration pressure, we assumed that it is a constant fraction of the increasing core-mantle boundary pressure.
This constant fraction was refined along with the initial volatile compositional gradients in the disk and the parameters associated with sulfide segregation by a least squares minimization so that the Earth-like planet's mantle matches that of the bulk silicate Earth once planetary accretion and differentiation are complete.
The values of the fitted parameters are listed in the supplementary information.
The Venus-like planet evolved according to the same parameters as the Earth-like planet.

\begin{figure*}
\centering
\includegraphics[width=\textwidth]{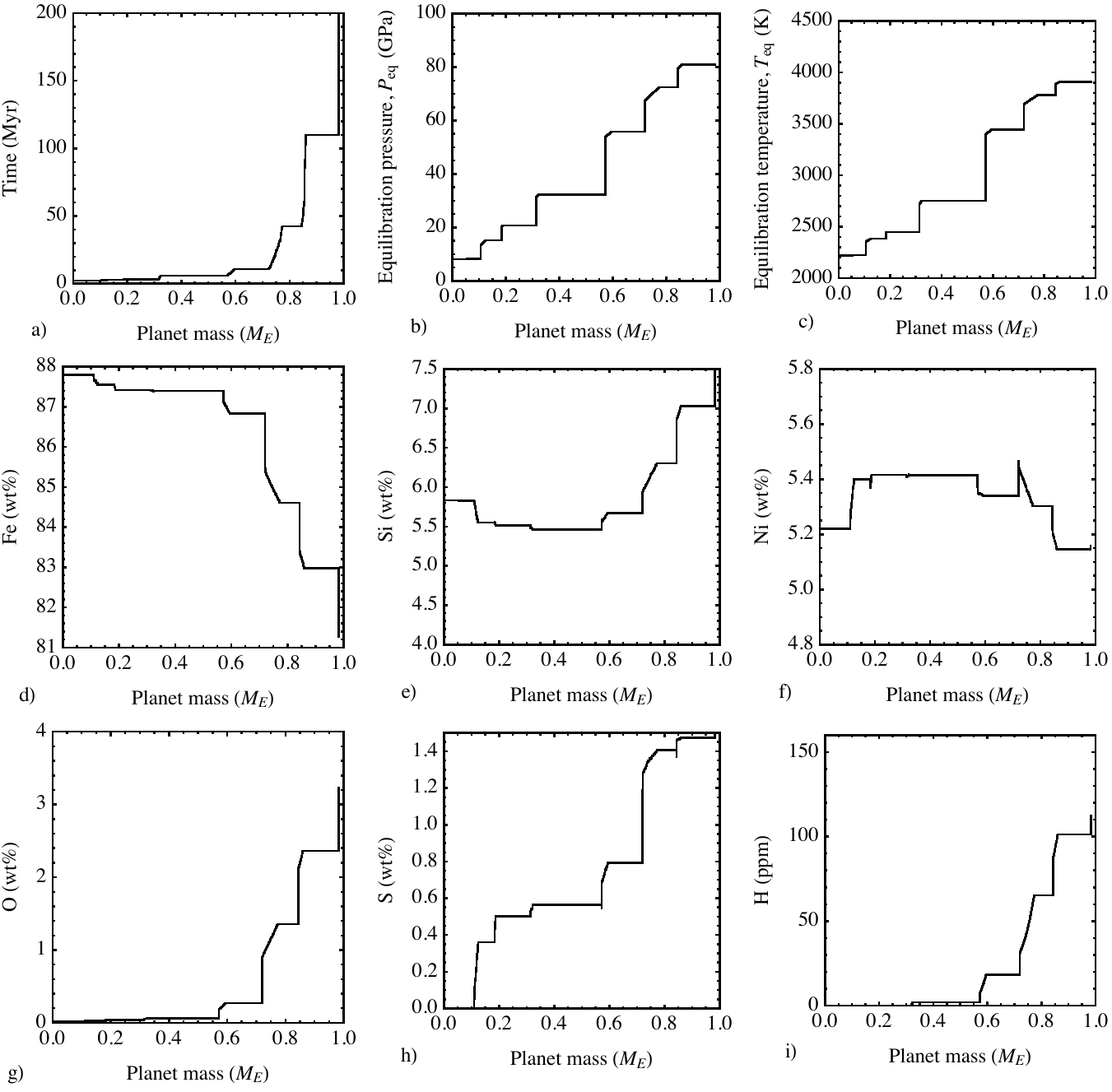}
\caption{The growing mass (panel a), metal-silicate equilibration pressure (panel b), temperature (panel c), and the evolving bulk core composition (panels d--h) as a function of accreted mass for the Earth-like planet.}
\label{fig:EarthGrowth}
\end{figure*}

The composition of each core addition changes significantly during the course of accretion as the equilibration pressures and temperatures change, see Fig.~\ref{fig:EarthGrowth}.
The final bulk composition of the core and mantle of the Earth-like planet is shown compared to the estimated composition of the bulk silicate Earth from \citet{Palme:2003dp} in the supplementary information.
The light elements Si and O are increasingly incorporated into the core, since they increasingly partition into the metal as pressures and temperatures increase \citep{Frost:2010gya,Mann:2009jj}.
While sulfide segregation does add a little sulfur to the core ($\sim$ 0.1~wt\% total), the vast majority is added during normal core formation and it also preferentially partitions in to the core at higher pressures and temperatures \citep{Boujibar:2014bd}.
Our primitive model for hydrogen partitioning adds H to the core forming liquids whenever water is disassociated during metal-silicate equilibration \citep{Rubie:2015fj}; this happens rarely and only small trace amounts of H are partitioned into the core.
Potential other light elements in the core such as carbon are not considered, however it is unlikely that they would partition into the core in such quantities \citep[see Table~1 of][]{Hirose:2013io} and in such a pattern as to cancel the effects of Si, O and S.

The increasing abundance of light elements in the core as a function of time is recorded in the radial stratigraphy of the core, see Fig.~\ref{fig:asaccreted}. 
We modeled the effect of changing radial composition on the core density as a perturbation to the reference density $\rho_\text{ref}$ given by the Murnaghan equation of state in Eq.~\ref{eqn:coredensity}.
The combined thermal and compositional perturbations to the density $\rho$ of a core layer are:

\begin{equation}
\rho = \rho_\text{ref} \left[ 1 - \alpha \left( T - T_\text{ref} \right) - \sum_i \beta_i \left( C_i - C_{i,\text{ref}} \right)  \right]
\end{equation}

\noindent where for each element $i$ (Si, O, S and H) in the core, $C_i$ is the molar composition of the core layer, $C_{i,\text{ref}}$ is the bulk molar composition of the core, and $\beta_i$ is the estimated compositional expansivity for each element in liquid iron at core pressures and temperatures: 0.32 for O, 0.45 for Si, 0.38 for S, and 0.29 for H \citep{Alfe:2002ke,Antonov:2002jo}.

In general, a stable density stratification is established, since the core additions deposited last are at the top of the core stratigraphy, equilibrated at the highest pressures and temperatures, and have the highest light element abundances and lowest densities.
In detail, the composition of each new core addition depends on the composition of the incoming projectile core and the interacting portion of the target's mantle as well as the increasing pressure and temperature of metal-silicate equilibration during accretion.
The projectile material varies in composition because of the heliocentric radial mixing observed in protoplanetary disk dynamics \citep{Chambers:2001kt}.
Thus, the accreted composition profile is not a smooth function of radius but reflects a stratigraphy established by multistage core formation.

\begin{figure}
\centering
\includegraphics[width=9cm]{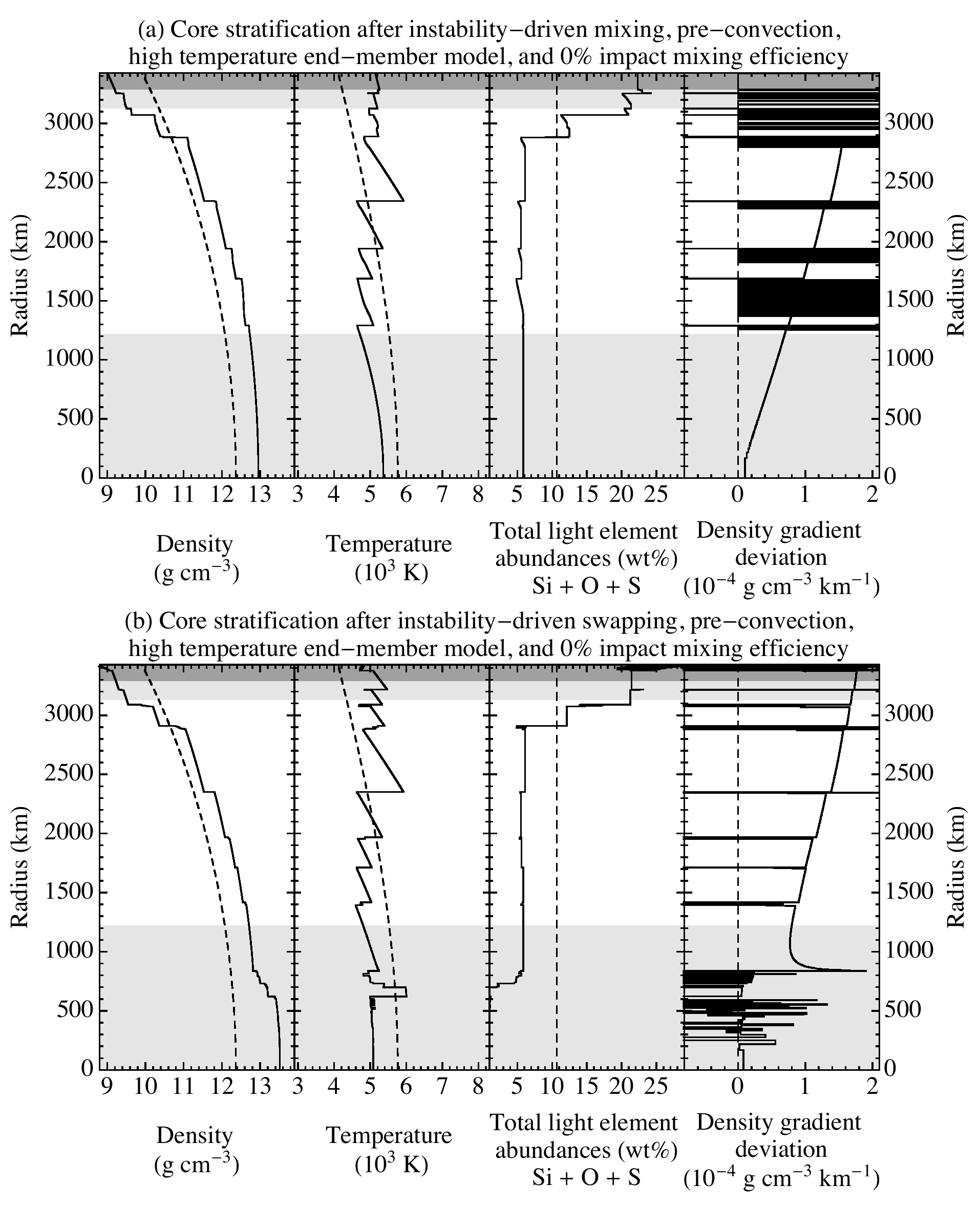}
\caption{As solid lines, the panels show the radial profiles of the density ($\rho$; left-most panel), temperature ($T$; left-center panel), wt\% abundance of total light elements (Si, O and S; right-center panel), and the deviation of the density gradient from an isentrope ($\partial \rho /\partial r - \left. \partial \rho / \partial r \right|_S$; right-most panel) of the core of the Earth-like planet.
As explained in the text, even small light element abundance gradients which are difficult to discern in the plot can generate negative density gradient deviations.
Both subfigures show the final core profile from models that include density stabilization and convective mixing after every accretion event, however each shows a different density stabilization models: (a) mixing and (b) swapping, as described in the text.
These models all use the high temperature thermal end-model, but there is no significant difference with the other thermal models.
The reference model shown as a dashed line, and gray regions are the same as that described in Fig.~\ref{fig:asaccreted}.
}
\label{fig:afterdensity}
\end{figure}

Some core additions have lower light element abundances than those beneath them, which is in contrast to the general trend established by metal-silicate equilibration.
Indeed, these core additions are density unstable (see Fig.~\ref{fig:asaccreted}), i.e. more dense layers are above less dense layers so that $\partial \rho / \partial r > 0$.
It is not clear how the core relaxes during accretion to achieve a stable density configuration, so we implemented two different density stabilization models: mixing and swapping.
A layer mixing model is appropriate if density instabilities during the core addition process result in turbulent flows, while a layer swapping model is appropriate if the flows are laminar.
For the mixing model, when a core layer is more dense than the layer immediately below it, the two layers fully mix creating two new layers with identical compositions and potential temperatures.
If these new layers are again more dense than the layers beneath them, this process continues until the core has a stable density stratification.
For the swapping model, layers exchange position rather than mix until stability is reached, adiabatically cooling and heating as appropriate. 
Reality lies between these two end-member models.
Both are presented in Fig.~\ref{fig:afterdensity}, and overall, the two models give similar results preserving the stable stratification observed in Fig.~\ref{fig:asaccreted}.
The swapping model preserves more density structure than the mixing model, since the record of every core accretion event is preserved, but just not in the order of accretion.

\subsection{Effect of convection on the structure of the core}
Density stabilization models, described above, are distinct from possible mixing due to convection, which occurs when the Rayleigh number Ra of the core fluid exceeds a critical value Ra$_c$ and instabilities are driven either by a thermal gradient in the case of thermal convection or combined thermal-compositional gradients in the case of double diffusive convection, so:

\begin{equation}
\text{Ra}_c < \text{Ra} = \frac{g L^4  }{ \mu \kappa } \left(\frac{\partial\rho}{\partial r} - \left.\frac{\partial\rho}{\partial r}\right|_{S}\right)
\end{equation}

\noindent where $g$ is the local gravitational acceleration, $L$ is a length scale, $\mu$ is the dynamic viscosity, $\kappa$ is the thermal diffusivity, $\partial \rho / \partial r$ is the local density gradient, and $\left. \partial \rho / \partial r\right|_S$ is the isentropic density gradient \citep{Kono:2001ik}.
When considering the Rayleigh number Ra, order of magnitude estimates for the gravitational acceleration ($g = 5$~m~s$^{-2}$), viscosity \citep[$\mu = 1.25 \times 10^{-2}$~Pa~s;][]{deWijs:1998bh}, and thermal diffusivity \citep[$\kappa = 2 \times 10^{-5}$~m$^2$~s$^{-1}$;][]{Pozzo:2012ik} combined with a layer width $L \gtrsim 1$~km produce a very large positive value for the term in front of the density gradient deviation: $g L^4 / \mu \kappa \gtrsim 2 \times 10^{19}$~g$^{-1}$~cm$^{3}$~km.

The critical Rayleigh number for Earth's outer core is estimated to be about $\text{Ra}_c \sim 10^{16}$ \citep{Gubbins:2001jb} and extends up to $10^{17}$ when considering the entire core.
Changing the layer location and its thickness changes the Ekman number \citep{Kono:2001ik} and the shell geometry \citep{AlShamali:2004ir}, both of which effect the critical Rayleigh number, so the range of possible layer critical Rayleigh numbers extends down to about $10^{9}$ when considering a km thick region near the center of Earth.
Given that the critical Rayleigh number is smaller than the pre-factor $g L^4 / \mu \kappa$ from Eq. 12 by a factor of about $10^{-11}$~g~cm$^{-3}$~km$^{-1}$ at most, convection is expected if the density gradient deviation has a positive value, so:

\begin{equation}
0 < \frac{\partial\rho}{\partial r} - \left.\frac{\partial\rho}{\partial r}\right|_{S} 
\end{equation}

\noindent Conversely, if the density gradient deviation is negative, then the core cannot convect locally and heat must be transported via conduction.

The density gradient deviation is the density gradient relative to that established by a constant composition isentrope, so it is a sum of thermal and compositional terms: 

\begin{equation}
\frac{\partial\rho}{\partial r} - \left.\frac{\partial\rho}{\partial r}\right|_{S} = - \rho_\text{ref} \left[ \alpha \left(\frac{\partial T}{\partial r} - \left.\frac{\partial T}{\partial r}\right|_S \right) + \sum_i \frac{\beta_i}{\tau} \frac{\partial C_i}{\partial r} \right]
\end{equation}

\noindent where the thermal term contains the local temperature gradient $\partial T/\partial r$ and the adiabatic temperature gradient $\left.\partial T/ \partial r \right|_S$, i.e. the local gradient of the reference thermal model defined in Eq.~10, and where the compositional term sums over the local chemical gradient $\partial C_i / \partial r$ of each of the considered light elements ($i=$ Si, O, S and H).
We assume that the chemical diffusivity $\kappa_c = 5 \times 10^{-9}$ m$^2$ s$^{-1}$ is the same for all considered elements \citep{Posner:2017ia,Posner:2017gj}, and so the compositional structure established by multistage core formation will not be homogenized by diffusion, even over the age of the solar system.
Lastly, the ratio of chemical to thermal diffusivities $\tau = \kappa_C / \kappa = 2.5 \times 10^{-4}$.

Density gradient deviation profiles of the newly formed core (see Fig.~3) have two components: (1) a steep trend line and (2) nearly horizontal departures from that trend.
The steep trend line is established by the retention of gravitational potential energy during core formation within the core addition.
When all of the dissipated gravitational potential energy is retained in the layer as heat (high temperature end-member model), the steep trend line in the mixing layer model is approximately $6 \times 10^{-8} \left(r / \text{km} \right)$~g~cm$^{-3}$~km$^{-1}$, where $r$ is the radius.
As the fraction of retained heat from dissipated gravitationally potential energy goes to zero (the low temperature or adiabatic end-member models), the slope of the trend line goes to infinity and the value of the density gradient deviation goes to zero.
Given that the critical Rayleigh number is effectively zero, practically any heat retention will produce a core with layers as predisposed towards thermal convection as the high temperature end-member model, even though the exact Rayleigh number may be many orders of magnitude different.
However, the thermal convection will be bounded between the horizontal negative density gradient deviations.

\begin{figure}
\centering
\includegraphics[width=9cm]{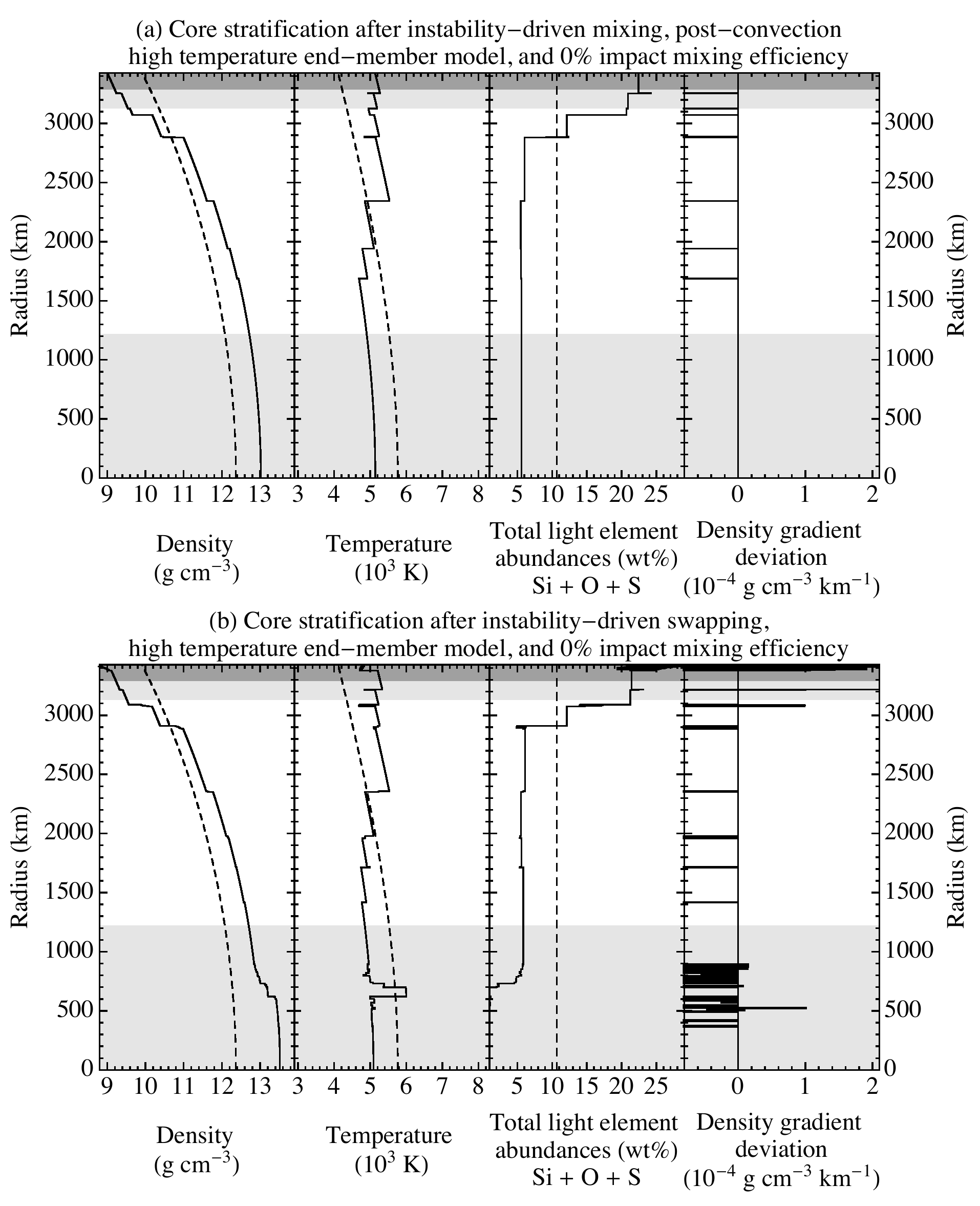}
\caption{As solid lines, the panels show the radial profiles of the density ($\rho$; left-most panel), temperature ($T$; left-center panel), wt\% abundance of total light elements (Si, O and S; right-center panel), and the deviation of the density gradient from an isentrope ($\partial \rho /\partial r - \left. \partial \rho / \partial r \right|_S$; right-most panel) of the core of the Earth-like planet.
As explained in the text, even small light element abundance gradients which are difficult to discern in the plot can generate negative density gradient deviations.
These subfigures are identical to those in Fig.~\ref{fig:afterdensity} except that convection has mixed the contiguous positive density gradient deviations regions in Fig.~\ref{fig:afterdensity}, so those same regions appear generally in this figure to have zero density gradient deviations.
There are a few exceptions in the swapping subfigure (b) where individual layers with positive density gradient deviations at the resolution limit of the model have not completely relaxed to zero.
These models all use the high temperature thermal end-model, but there is no significant difference with the other thermal models.
The reference model shown as a dashed line, and gray regions are the same as that described in Fig.~\ref{fig:asaccreted}.
}
\label{fig:afterconvection}
\end{figure}

Horizontal departures from this vertical trend line are imposed by compositional and thermal shifts associated with different core formation events.
Negative density gradient deviations are impermeable conductive barriers to convection, since they are the locations of boundaries between core additions with different light element abundances.
At these boundaries, the light element abundance gradient is typically positive $\partial C_i / \partial r \gtrsim 10$~molar~ppm~km$^{-1}$.
Re-arranging Eqs.~13 and 14 and estimating that the adiabatic radial temperature gradient is $\partial T / \partial r |_{S} \sim - 0.5$~K~km$^{-1}$, the Rayleigh criterion for convection requires a strong negative temperature gradient: $\partial T / \partial r \lesssim - 1000$~K~km$^{-1}$.
These conductive layers prevent the general homogenization of the core's composition by prohibiting neighboring regions, which may experience convection independently, from compositionally equilibrating with each other.
Thus, these barriers generally persist even after thermal convection due to core cooling has mixed contiguous regions with positive density gradient deviations, as shown in Fig.~\ref{fig:afterconvection}.
In effect, they create an onion-like shell structure within the core, where convective mixing eventually homogenizes the fluids within each shell but prevents homogenization between shells.

With respect to double diffusive instability theory, the density stratification established by multistage core formation is of the `diffusive' type since colder, lower mean molecular weight liquid is above hotter, higher mean molecular weight liquid and the diffusivity ratio $\tau < 1$.
Unlike the `fingering' type double diffusive instability, in this regime a perturbation grows via oscillations overshooting a background stable stratification increasing diffusion rates across the boundary.
The criterion for this instability is that the density ratio $R_\rho = \beta (\partial C / \partial r ) / \alpha \left( \partial T/\partial r - \left.\partial T/ \partial r \right|_S \right)$ is between $1 < R_\rho < R_{\rho,c}$ where the critical density ratio is $R_{\rho,c} = \left( \text{Pr} + 1 \right) / \left( \text{Pr} + \tau \right) \approx 20$ when the Prandtl number is $\text{Pr} = \mu / \kappa \rho \approx 0.05$.
This condition is never met, because the density ratio is always either $R_\rho \ll 1$ at the conductive boundary or $R_\rho \gg R_{\rho,c}$ within the homogenous layers.

\begin{figure}
\centering
\includegraphics[width=9cm]{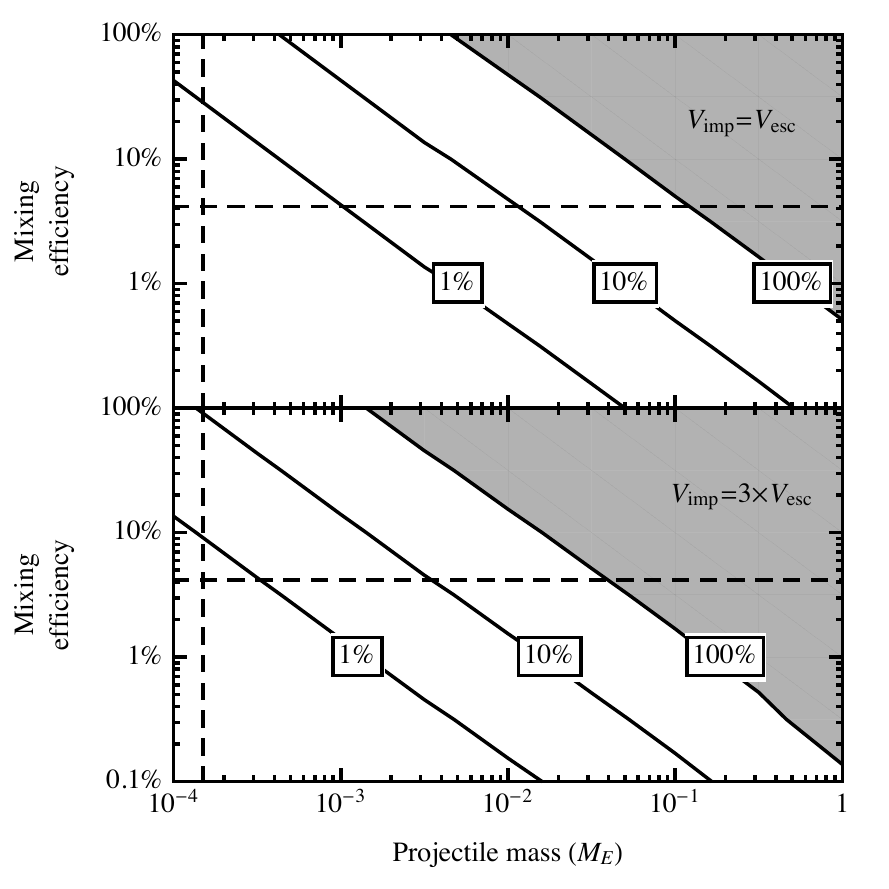}
\caption{Each panel shows contours (1\%, 10\%, and 100\%) of constant core mixing fraction as a function of the projectile mass in Earth masses and the mixing efficiency.
Remember that the mixing fraction describes the fraction of each core layer that is mixed amongst the layers while the mixing efficiency describes how much energy of the total released impact energy ultimately goes into mixing the core.  
These contours are generated for a hypothetical impact occurring on the Earth-like planet modeled in Fig.~\ref{fig:afterconvection} (a), so the target is an Earth-mass body with the stratified core shown there.
The projectile, which has its mass given by the abscissa and a core mass one-third of its total, impacts at a different impact velocity for each panel (the escape velocity in the upper panel and three-times the escape velocity in the lower panel).
The mixing efficiency along the ordinate dictates what fraction of the total released impact energy is converted into delivered core mixing energy.
The delivered core mixing energy divided by the energy required to mix the core completely is the mixing fraction, which is given by the contours.
Gray regions are where the entire core is entirely mixed, i.e. mixing fractions equal to or greater than one.
The nominal mixing efficiency of about 4\% is shown as a horizontal dashed line.
For reference, a projectile with the mass of Ceres, the largest asteroid, is shown as a vertical dashed line.
}
\label{fig:mixingcontours}
\end{figure}

\section{Consequences of impacts on core structure}
Energy released during impacts could turbulently mix the core, thus significantly altering its density structure \citep{Stevenson:2014ve} and erasing the stratigraphy produced by multistage core formation.
We simulate the role of impacts by first calculating the required energy to entirely mix the core prior to the impact, i.e. the potential energy difference between the current core and a completely mixed core with an adiabatic temperature profile \citep{Nakajima:2016ua}.
Second, we calculate the total energy released during the impact, both the kinetic energy of the impact and also the potential energy difference as the two differentiated bodies merge into a single differentiated body.
We determine these quantities directly from the N-body simulation and the calculated interior structures of the pre- and post-impact bodies.
In order to determine the delivered mixing energy, we multiply the total released energy by a mixing efficiency, which we estimate to zeroth order to be about 4\% as described below.
Lastly, the delivered mixing energy is divided by the required mixing energy from the first step to obtain the mixing fraction.
Examples of this calculation are shown in Fig.~\ref{fig:mixingcontours}.
If the mixing fraction is at, or exceeds 100\%, then the core is completely mixed.
Otherwise, only a fraction of the core is mixed.

In the case of a mixing fraction less than 100\%, we use a first order mixing model that divides each core layer into two parts corresponding to a mixed fraction and a preserved fraction.
The mixed fraction of every layer is mixed together across the entire core, completely thermally and compositionally equilibrating, as if violently stirred during the impact.
Then the mixed fraction of each layer is recombined with the preserved fraction of that layer, recognizing that lateral mixing is ultimately more vigorous and continues longer than radial mixing post-impact since it does not have to do work against the gravitational potential.
In this way, the amplitude of the thermal deviations of the temperature profile away from an isentrope and compositional deviations away from a homogenous bulk composition decrease proportionally to the mixing fraction.

Despite the simplicity of estimating the total released energy, it is difficult to estimate the delivered mixing energy.
From laboratory experiments, we know that only about a quarter of the kinetic energy deposited in fluid motion is expected to be dissipated into turbulent mixing \citep{Mcewan:1983gh}.
But we do not expect all released impact energy to be deposited into the core as fluid motion, instead much of the released impact energy goes into other reservoirs such as the mantle, atmosphere and impact ejecta, and other processes such as heating, large-scale melting, and bulk rotation.
Furthermore, the energy will be deposited heterogeneously with much of it concentrated in the mantle directly beneath the impact site and at the antipode.
Determining the fraction of the total energy released in an impact that is eventually used to mix the core is a complicated task left for future work since it likely will require significant laboratory and numerical experiments.

Instead, we have created a quantity called the mixing efficiency, which converts the total released impact energy into the core mixing energy.
For instance, if the mixing efficiency is 0\%, then none of the energy released during the impact produces core mixing and the results are those shown in Fig.~\ref{fig:afterconvection}.
However, if the mixing efficiency is 1\% and the total released impact energy is five times the required mixing energy, then the mixing fraction is 5\%. 
A zeroth order estimate for the mixing efficiency may be obtained by assuming that the impact energy is equally distributed amongst all mass elements so the core receives one-third, that the impact energy delivered to the core is equally partitioned between internal and kinetic energy, and that the turbulent mixing energy is a quarter of the kinetic energy deposited in the core.
From these assumptions, the mixing efficiency is about $1/3 \times 1/2 \times 1/4 = 4$\% with an approximate order of magnitude of uncertainty.

The cores of relatively large differentiated projectiles (like that in a Moon-forming impact; core radius $\gtrsim 500$~km) merge rapidly with the target core on the same timescale as dynamical relaxation of the newly merged planet~\citep[$\tau_\text{merge} \sim \tau_{relax} \sim $ hours;][]{Dahl:2010ik,Cuk:2012hj,Canup:2017us}.
However, metal from either differentiated \citep[core radius $\lesssim 500$~km;][]{Deguen:2014dq} or undifferentiated planetesimals is first turbulently mixed into the target mantle before sinking much more slowly to the core-mantle boundary \citep[$\tau_\text{merge} \gtrsim 1000$ hours $\gg \tau_{relax} \sim $ hours;][]{Dahl:2010ik,Rubie:2016kx}.
Thus, core forming liquids from small projectiles are not mechanically mixed into the core by their own impact unlike core material from large projectiles. 

However, small projectiles do not mix the core.
In Fig.~\ref{fig:mixingcontours}, we show an example suite of core mixing calculations for a hypothetical projectile striking the fully grown Earth-like planet shown in Fig.~\ref{fig:afterconvection} (a).
For projectiles near Ceres-size or smaller, they deliver on order of a per cent or less of the energy required to mix the core
Thus, even if the core mixing efficiency is high, they do not substantially contribute to core mixing.
On the other hand, giant impacts often deliver many times the energy necessary to entirely mix the core so the exact value of the mixing efficiency becomes important.
If the mixing efficiency is greater than the nominal value of 4\%, then a Mars-sized projectile (about a tenth of an Earth mass) like the canonical Moon-forming impactor \citep{Canup:2001te} is sufficient to completely mix the core.
Alternatively, if the mixing efficiency is less than the nominal value of 4\%, then only the largest projectiles will mix all or most of the core.

Recall that uncertainty of the nominal mixing efficiency is currently high, but if knowledge of this value improves, then it could discriminate between leading Moon-forming impact scenarios.
If the mixing efficiency is typically much lower than 4\%, then only the highest velocity impacts or those between like-sized bodies would result in mechanical mixing of the core, either of which sound remarkably like an already proposed Moon-forming impact scenario \citep{Cuk:2012hj,Canup:2012cd}.
However, the opposite could be true as well, and Earth's core could be well-mixed by a series of lunar-sized projectiles during its growth, consistent with a multiple impact Moon-forming scenario \citep{Rufu:2017ea}.

\begin{figure}
\centering
\includegraphics[width=9cm]{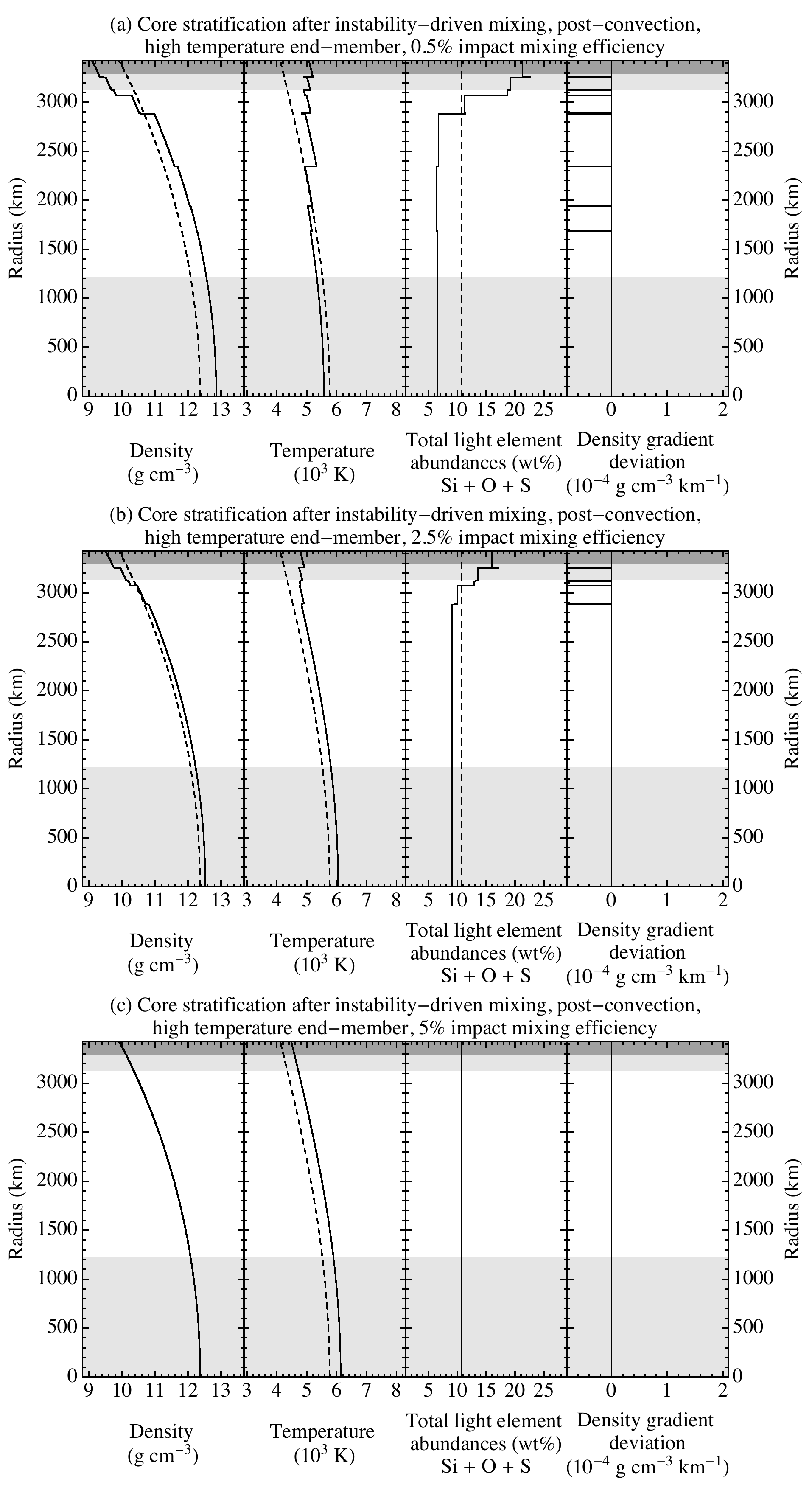}
\caption{As solid lines, the panels show the radial profiles of the density ($\rho$; left-most panel), temperature ($T$; left-center panel), wt\% abundance of total light elements (Si, O and S; right-center panel), and the deviation of the density gradient from an isentrope ($\partial \rho /\partial r - \left. \partial \rho / \partial r \right|_S$; right-most panel) of the core of the Earth-like planet.
As explained in the text, even small light element abundance gradients which are difficult to discern in the plot can generate negative density gradient deviations.
The Earth-like planet reference model, shown as a dashed line, and gray regions are the same as in Fig.~\ref{fig:asaccreted}.
All subfigures show the final core profile from models that include density stabilization via the mixing model, convective mixing, and mixing induced from impacts after every accretion event, however each shows a different mixing efficiency: (a) 0.5\%, (b) 2.5\%, and (c) 5\%, as defined in the text.
In other words, these models are identical to that shown in Fig.~\ref{fig:afterconvection} (a) plus the accumulated effect of impact driven core mixing.
These models all use the high temperature thermal end-model.
}
\label{fig:afterimpacts}
\end{figure}

Using the N-body accretion history which provides the characteristics of each impact including the impact velocity, we simulate impact mixing of the core for each impact throughout all of accretion.
For the sake of simplicity, we define a constant mixing efficiency throughout all of accretion for each model.
This is unlikely to be fully realistic, since the mixing efficiency is likely to vary with impact geometry and velocity, target rotation state, physical state of the target's mantle (solid or molten) etc., but the final structure of the core is mostly determined from the final giant impact or two.
We show the accumulated effects of this impact driven core mixing on the final core structure in Fig.~\ref{fig:afterimpacts} for mixing efficiencies of (a) 0.5\%, (b) 2.5\%, and (c) 5\% (the results in Fig.~\ref{fig:afterconvection} (a) are for a constant mixing efficiency of 0\%).
It is clear that as the mixing efficiency increases, the core becomes increasingly homogenous and the number of conductive barriers decreases from the bottom upwards.
At a 5\% mixing efficiency, the entire core is well mixed and homogenous.
Considering that Earth appears to possess a nearly homogenous core \citep{Mandea:2012hy}, the mixing efficiency for Earth must have been near this value, which is generally in good agreement with the 4\% estimate made earlier.
While none of these models result in a layer of conductive bands as narrow as the stratified layers observed by seismic waves \citep{Helffrich:2010cy} or magnetic fields \citep{Buffett:2014jd}, it's clear that a generalized version of this process could result in a distinct region of stratified lower density material in the uppermost layers of Earth's core when mixing efficiencies are a few per cent.

\section{Consequences of core structure on planetary magnetic fields}
If the mixing efficiency is high, then the core may convect as a single cell and a traditional planetary dynamo may result.
When the mixing efficiency is low, the core is divided into convective shells between the conductive boundaries created by multistage core formation and revealed as negative horizontal anomalies in the density gradient deviation profiles.
The convective shells between the conductive boundaries will meet the general requirements for sustaining a magnetic dynamo, as long as there is a source of buoyancy to drive the convection since the fluid is conductive and the Coriolis force is non-negligible even in the case of Venus \citep[as reviewed in][]{Stevenson:2003ji}.
Initially, any retained heat in the core from released gravitational potential energy can power convection in each shell as long as there is enough of a thermal gradient out of the top of the shell to drive the heat flux.
However, it is unclear whether the planetary magnetic field generated by this multi-celled onion-like convective structure will appear similar to Earth's current dipolar field.
Indeed, paleomagnetic records created on the planetary surface from a thin convecting shell are likely quite different than that recorded currently on Earth from it's thick convecting shell \citep{Stanley:2005bs}.
While convection is occurring in each shell of the core, there will also be interaction between the fields generated by each dynamo, but future work will be needed to fully determine the effects of this interaction.

As the core cools, the thermal gradient within each convective cell eventually relaxes to an adiabat and the density gradient deviation trend line relaxes to zero as seen in the low temperature and adiabat end-member models in Fig.~6.
Upper conductive boundaries, which all but the uppermost shell possesses, force heat to be transported out of the lower shell conductively, and thus heat will be transported through the upper shell conductively with no thermal convection.
In the uppermost shell, thermal convection is dependent on the core-mantle boundary heat flux, which could become large enough to initiate thermal convection in this shell only.

There are also other possible sources of buoyancy in the core such as the proposed magnesium precipitation at the top of the core \citep{Badro:2016es,ORourke:2016bw}, which like thermally driven convection could drive convection in only the uppermost shell in a planet with an onion-like core structure until that shell obtained a density as high as the shell beneath it, and so forth.
Alternatively, if inner core crystallization occurs then light elements are introduced into the outer core from the bottom creating a negative compositional gradient in the lowermost shell and driving convection in that shell only until it's density matches that of the shell above it, and so forth.
In such a case, a dynamo driven in the bottom-most shell may not be able to create a magnetic field detectable at the surface of the planet because of the large amounts of conductive fluid material between the convecting cell and the surface \citep{Christensen:2008kh}.

In general, the magnetic Reynold's number is the ratio of the rates of magnetic induction (i.e. fluid advection) to magnetic diffusion: $Re_M = v L / \lambda$, where $v$ is a characteristic velocity, $L$ is the convective layer thickness and $\lambda$ is the magnetic diffusivity of core liquids.
Below the critical magnetic Reynold's number ($Re_M \sim 10$), the magnetic field diffuses away faster than the dynamo action can generate it.
The estimated value of the magnetic Reynold's number for Earth's outer core is about $Re_M \lesssim 10^3$ \citep{Davidson:2013gx}.
If the impact mixing efficiency is low and numerous conductive boundaries exist, then the relevant length scale for such a planet would be approximately an order of magnitude smaller (perhaps, changing from $L \sim 2000$~km to $L \sim 200$~km), thus reducing the magnetic Reynold's number by an order of magnitude as well, even if the characteristic velocities remain unchanged.
A current leading theory for the lack of a planetary magnetic field on Venus is that the lack of plate tectonics has stifled heat flow through the mantle, thus through the CMB, and eliminating the power for core convection and the characteristic velocity $v$ of core fluids \citep{Nimmo:2002io}.
In the case that the core of Venus is divided into many layers by conductive boundaries due to multi-stage core formation, this condition on the mantle heat flow is relaxed.
If the layers are small enough, then the magnetic Reynold's number will be sub-critical directly due to a small convective domain.

\begin{figure*}
\center
\includegraphics[width=\textwidth]{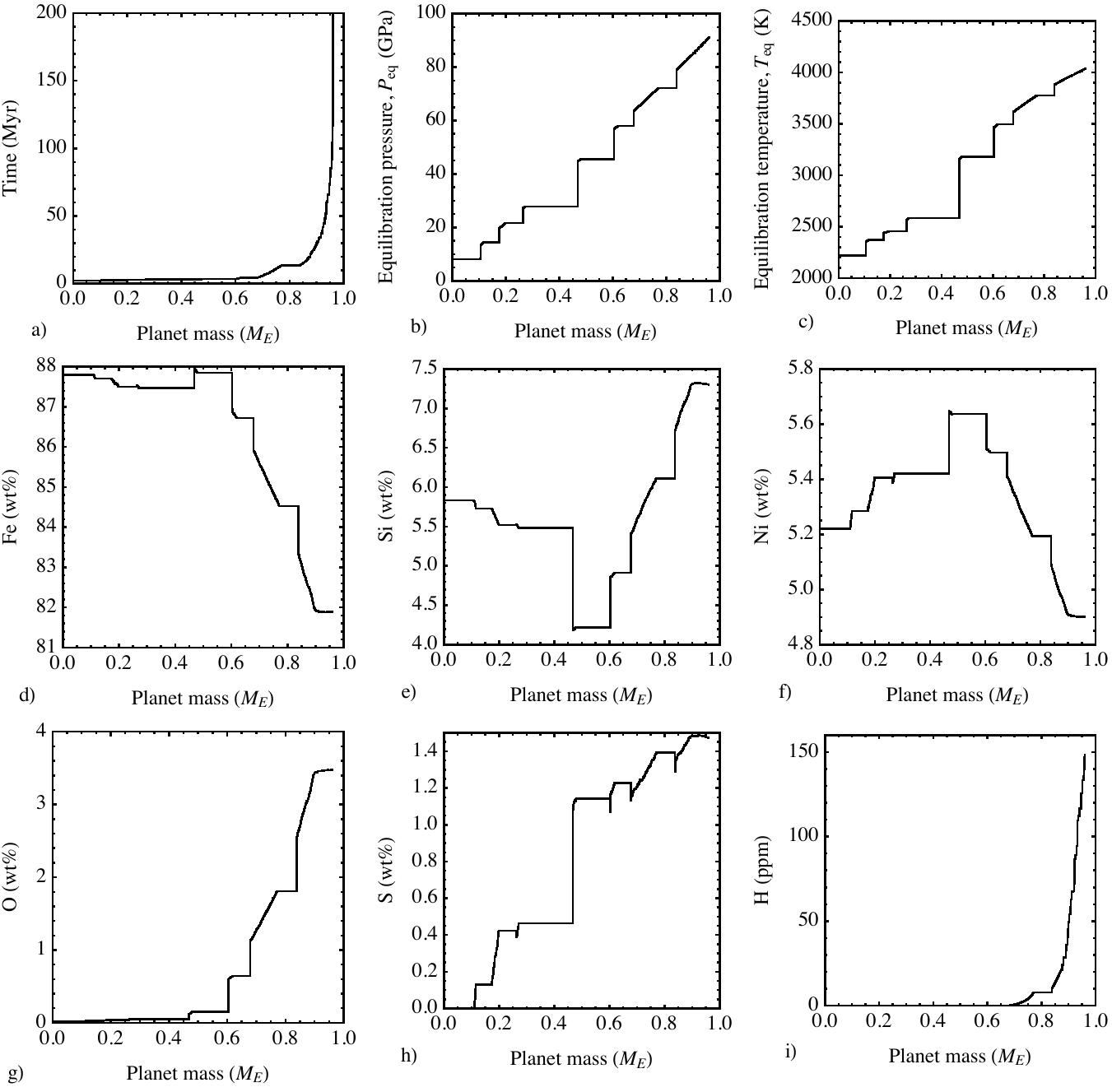}
\caption{The growing mass (panel a), metal-silicate equilibration pressure (panel b), temperature (panel c), and the evolving bulk core composition (panels d--h) as a function of accreted mass for the Venus-like planet.}
\label{fig:VenusGrowth}
\end{figure*}

\begin{figure}
\centering
\includegraphics[width=9cm]{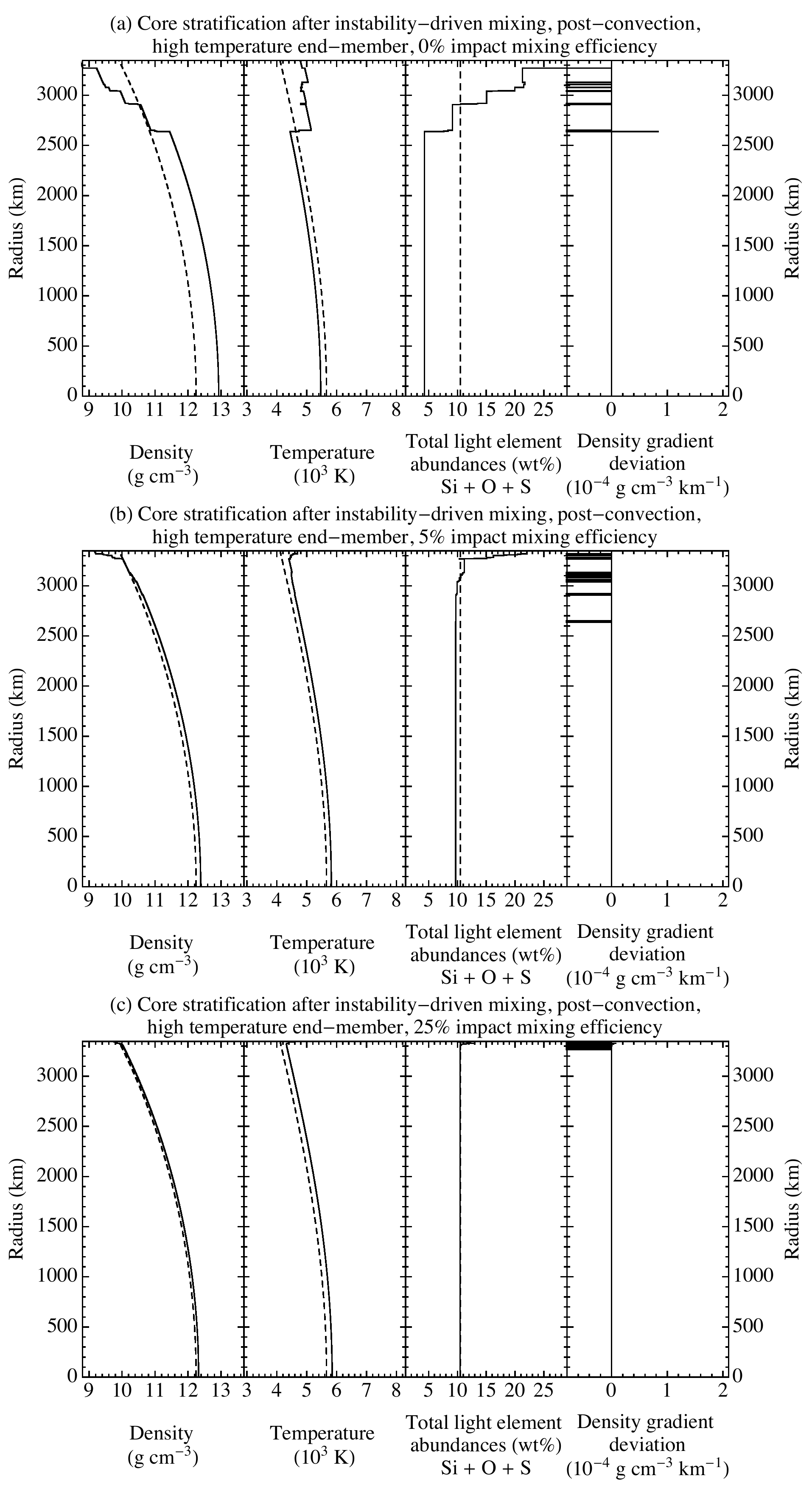}
\caption{As solid lines, the panels show the radial profiles of the density ($\rho$; left-most panel), temperature ($T$; left-center panel), wt\% abundance of total light elements (Si, O and S; right-center panel), and the deviation of the density gradient from an isentrope ($\partial \rho /\partial r - \left. \partial \rho / \partial r \right|_S$; right-most panel) of the core of the Venus-like planet.
As explained in the text, even small light element abundance gradients which are difficult to discern in the plot can generate negative density gradient deviations.
A completely-mixed adiabatic reference model built from a fit to the outer core of the preliminary reference Earth model \citep[PREM;][]{Dziewonski:1981bz} extrapolated to Venus and possessing a homogenous composition identical to that of the bulk Venus-like planet's core is shown as a dashed line.
All subfigures show the final core profile from models that include density stabilization via the mixing model, convective mixing, and mixing induced from impacts after every accretion event, however each shows a different mixing efficiency: (a) 0\%, (b) 5\%, and (c) 25\%, as defined in the text.
These models all use the high temperature thermal end-model.
}
\label{fig:venus}
\end{figure}

\section{Core structure of Earth and Venus}
The Moon-forming impact on Earth was a late \citep{Jacobson:2014cm} and violent \citep{Cuk:2012hj,Canup:2012cd} event that likely delivered enough mixing energy to the core to remove all traces of multistage core formation \citep{Nakajima:2016ua}.
Thus Earth's core likely appeared as in Figure~\ref{fig:afterimpacts} (c) soon after the end of core formation.
However, we hypothesize that this may not be the case for Venus, since it has no detectable internally-driven planetary magnetic field despite many similarities with Earth.
To test this idea, we performed an identical analysis for the Venus-like planet which grew in the exact same N-body simulation as the Earth-like planet.
The details of its accretion, core formation history and light element abundances are shown in Fig.~\ref{fig:VenusGrowth} and its final bulk composition is tabulated in the supplementary information.
In Fig.~\ref{fig:venus}, we show the final core structures for three models with different mixing efficiencies: (a) 0\%, which can be directly compared to the Earth-like planet's core structure in Fig.~\ref{fig:afterconvection} (a), (b) 5\%, which can be directly compared to the Earth-like planet's core structure in Fig.~\ref{fig:afterimpacts} (c), and (c) 25\%.
The most striking difference between the two planets is that at a mixing efficiency of 5\%, the conductive layering in the core of the Earth-like planet has been eliminated whereas it still exists throughout the outer half of the mass (radius of about 2650 km) of the Venus-like planet's core.

\begin{figure}
\centering
\includegraphics[width=\columnwidth]{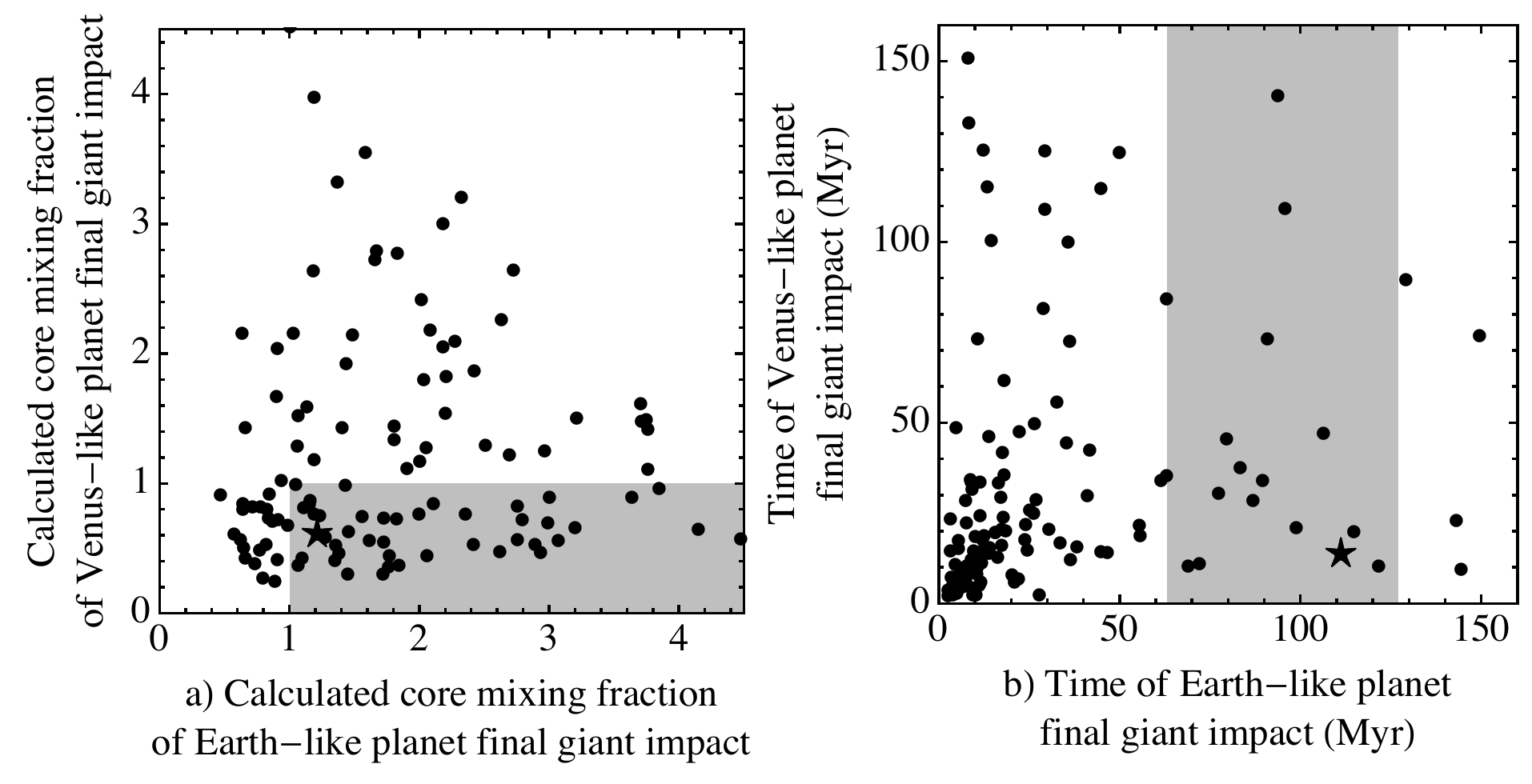}
\caption{Panel (a) shows the calculated core mixing fractions for pairs of Earth-like and Venus-like planets from 127 previously published planet formation simulations \citep{Walsh:2011co,Jacobson:2014it, Jacobson:2014cm}.
Panel (b) shows the times relative to the formation of the first solids in the solar system (CAIs) of the last giant impacts on the Earth-like and Venus-like planets of the same 127 simulated solar systems.
In all of these systems, only two terrestrial planets are created within the mass ranges (factor of two) of Earth and Venus and the planets modeled in detail throughout the paper are marked with a star.
In panel (a), the calculated core mixing fraction is the mixing efficiency, which is assumed to be 4\%, multiplied with the total released impact energy, which is determined from the planet formation simulations, and divided by the energy necessary to mix the core, which is determined from the core density profile shown in Fig.~\ref{fig:afterconvection}~(a) for Earth-like planets and Fig.~\ref{fig:venus}~(a) for Venus-like planets.
For illustrative purposes, we have not placed a ceiling on the calculated core mixing fraction at one, but in reality a core mixing fraction greater than one is equivalent to one, since the core is mixed at one regardless of the extra mixing.
The gray region indicates the region where the Earth-like planet experiences complete core mixing during the final impact and the Venus-like planet does not.
In Panel (b), the gray region indicates the one sigma bounds about the best estimate for the time of the last giant impact on Earth: $95 \pm 32$~Myr \citep{Jacobson:2014cm}.
The median time for the last giant impact on the Venus-like planet for these simulated Earth-like planets is $34$~Myr and 76\% have impacts prior to 63~Myr.
}
\label{fig:timing}
\end{figure}

This is a direct result of the differing impact histories between the two terrestrial planets.
From the highly siderophile element record on Earth, we know that Earth must have had a late ($\sim$95$\pm$32 Myr) giant impact but this constraint does not exist for Venus \citep{Jacobson:2014cm}.
In the simulation examined in this paper, the Venus-like planet has an early final giant impact ($\sim$11 Myr) at relatively low energy, while the Earth-like planet had a late giant impact ($\sim$110 Myr) at high energy (see Fig.~\ref{fig:timing}).
In fact, in 39\% of 127 previously published simulated planetary systems \citep{Walsh:2011co,Jacobson:2014it, Jacobson:2014cm}, the final impact on the Venus-like planet does not fully mix the core while the final impact on the Earth-like planet does.
This frequency increases to 50\% of the subset of systems, when we consider that we have independent evidence that the final Moon-forming impact was a high energy impact, so we remove from the denominator all systems with Earth-like planets that do not have fully mixed cores.
Particularly violent or gentle impacts can result at any time within the chaotic dynamics of planet formation as shown in Fig.~\ref{fig:timing}.
We can understand this result by examining again Fig.~\ref{fig:mixingcontours} and noting that increasing the impact velocity lowers the mixing efficiency required to completely mix the core by increasing the total released impact energy.
So two planets of similar size in the same protoplanetary disk can have very different core mixing histories, if they are the targets of projectiles with different masses and impact velocities.

Based again on Fig.~\ref{fig:venus}, the core of the Venus-like planet does not become completely mixed even in the case of 25\% mixing efficiency.
This is because while Earth and the Earth-like planet from the simulation have a late accreted mass of about 0.5\% of an Earth mass, constrained by the concentrations of highly siderophile elements in Earth's mantle \citep{Chou:1978uu}, the Venus-like planet has a much larger (13\% Earth masses) late accreted mass---the late accreted mass being the total mass delivered by planetesimals after that last giant impact.
While the mantle magma ocean created by the last giant impact exists, planetesimal accretion still contributes core forming liquids; afterwards, without giant impacts to create large quantities of melt, planetesimal cores are mixed by solid convection and oxidized in the mantle instead of being segregated significantly to the core \citep{Rubie:2016hl}.
While the mantle magma ocean created by the last giant impact is expected to exist for a few million years on each body, the two bodies accrete very different amounts of planetesimal mass during this period due to the differences in accretion rate between the Venus-like and Earth-like planet (the last giant impact on the Venus-like body occurs at 11~Myr while that on the Earth-like body occurs at 110~Myr).

The Earth-like and Venus-like planets highlighted in this paper, while representative of general trends, are not intended to be viewed as specific histories of Earth and Venus.
Instead, they are meant as archetypes. 
Both planets show the growth of concentric shell structures in their cores due to the changing conditions of metal-silicate equilibration during multi-stage core formation.
The radial location of conductive boundaries in the core and whether these boundaries persist against convection or impact-driven mechanical mixing is, in part, stochastic given the origin and nature of large projectiles during the era of terrestrial planet formation.
Thus, since it appears that multi-stage core formation is a natural outcome of terrestrial planet formation, we propose a hypothesis that the existence of a planetary magnetic field on Earth and the lack of a detectable field on Venus is due to distinct differences in the bombardment histories of these two sibling planets.
Namely, Venus avoided a large, violent impact near the end of its accretion, whereas the Earth was struck violently at the end of its growth, simultaneously creating its Moon and homogenizing its core.

\section{Conclusions}
In this paper, we modeled the accretion and differentiation of Earth-like and Venus-like planets from a collection of planetesimals and embryos in a protoplanetary disk to their final states.
In particular, we determined the composition of each core addition following metal-silicate equilibration and we also tracked the evolving physical and chemical state of the core with a number of end-member models, although no particular choice significantly effects the outcome with respect to core stratification.
We discovered that the cores of both planets grow stably stratified.
A stratigraphic record of the details of their accretion is maintained with upper layers containing higher abundances of light elements because they equilibrated at higher pressures and temperatures during their descent through the mantle.
These changes in composition create conductive bands throughout the core establishing a series of non-interacting convective shells.
While a thermally driven magnetic dynamo may be active in these shells initially, after subsequent cooling and the start of core solidification, a dynamo may only be driven in the lower-most shell.
This stratigraphic record in the core is maintained despite occasionally over-dense layers and thermal transport through the core, but it can be partly or completely destroyed by giant impacts.
The specific impact history of the planet, including how energetic and how efficiently impact energy is converted into mixing the core, matter greatly in determining whether the conductive barriers survive accretion.
Thus, the violence of accretion could separate those planets with planetary dynamos from those without.

\paragraph{Acknowledgments}
S.A.J., D.C.R. and A.M. were supported by the European Research Council (ERC) Advanced Grant ``ACCRETE'' [contract number 290568].

\clearpage
\section*{Supplementary Information}

\subsection*{Details of N-body simulations}
In order to understand the growth of Earth's core, we used previously published simulations of the growth of Earth from the accumulation of planetesimals and planetary embryos out of the terrestrial protoplanetary disk \citep{Jacobson:2014it}.
These simulations follow the now standard N-body approach, they used a symplectic N-body method \citep{Wisdom:1991ff} as modified in Symba to include perfect collisions \citep{Duncan:1998gn}.
For clarity, we focus on the results of a well-studied simulation, 4:1-0.5-8, which is the same as that examined in \citet{Rubie:2015fj,Rubie:2016hl}.
This simulation follows the prescription set out in the supplementary material of \citet{Walsh:2011co} in the section ``Saturn's core growing in the 2:3 resonance with Jupiter'' to model a Grand Tack terrestrial planet formation scenario.
In a self-consistent framework, the Grand Tack scenario reproduces the orbits and sizes of the terrestrial planets, particularly the small mass of Mars, and the compositional dichotomy of the asteroid belt.
However, the core formation results presented here do not depend on this scenario choice, since the terrestrial planets grow from a series of planetesimal and planetary embryo accretion events in all proposed terrestrial planet formation scenarios \citep{Raymond:2009is,Izidoro:2015bf}.

The simulated protoplanetary disk begins after the epoch of runaway growth during the epoch of oligarchic growth, so mass is bi-modally distributed between 87 planetary embryos each with half a Mars mass located between 0.7 and 3~AU and, out of computational necessity, only 2836 planetesimals with individual masses of $3.8 \times 10^{-4}$ Earth masses located between 0.7 and 3~AU and between 6 and 9.5~AU.
Such a bi-modal mass distribution in a protoplanetary disk can be created by either planetesimal accretion or pebble accretion \citep[as reviewed in][]{Jacobson:2015fn}.
The outer disk (6 to 9.5~AU) of planetesimals is dynamically scattered into the terrestrial planet forming region by the outward migration of the giant planets during the end of the Grand Tack scenario to deliver water to Earth and the C-complex asteroids to the main belt, whereas the region between 3 and 6 AU is initially occupied by the giant planets and contains no terrestrial planet building blocks \citep{Walsh:2011co}.
As the simulation proceeds, the protoplanetary disk self-stirs and enters the epoch of giant impacts during which the planetary embryos grow into the terrestrial planets from the accretion of planetesimals and other embryos.
The final system of planets in 4:1-0.5-8 consists of an Earth-like planet with a mass of 0.94 Earth masses at 0.97~AU, a Venus-like planet with a mass of 0.92 Earth masses at 0.62~AU, and two Mars-like planets with masses of 0.06 and 0.13 Earth masses at 1.67 and 1.79~AU, respectively.
This planetary system was selected from a suite of published simulations \citep{Jacobson:2014it} because the Earth-like planet had a late final giant impact and only accreted 0.5\% of its mass after that giant impact, which is consistent with the highly siderophile element record in Earth's mantle and other constraints \citep{Jacobson:2014cm}.

\subsection*{Planetary differentiation model parameters}

Table~1 contains the details of the important fitted parameters and their values for the planetary differentiation model, as described in detail in \citet{Rubie:2015fj} and \citet{Rubie:2016hl}.

\subsection*{Planetary bulk compositions}

Table~2 contains the final bulk compositions of the Earth-like and Venus-like planets.

\begin{table}
\begin{center}
\begin{tabular}{@{}ll@{}}
\toprule
\multicolumn{2}{c}{Best fit parameters} \\
\midrule
$X_{Si}^{met}$(1) & 0.106 \\
$X_{Fe}^{met}$(2) & 0.418 \\
$\delta_S$(0) & 0.864 AU \\
$\delta$(1) & 1.22 AU \\
$\delta$(2) & 1.59 AU \\
$\delta$(3) & 5.75 AU \\
$f_p$ & 0.68 \\ 
$f_m$ & 0.60 \\
$\tau_m$ & 6.08 Myr \\
\bottomrule
\end{tabular}
\end{center}
\caption{Important parameters and reduced chi squared for the planetary differentiation model \citep[for comparisons with other simulations see Table 4 of][]{Rubie:2015fj}.
The volatile (O, S, and H$_2$O) compositions of each body are set by heliocentric gradients in the protoplanetary disk with a number of fitted and fixed parameters.
$X_{Si}^{met}$(1) is the fraction of Si as metal interior to distance $\delta$(1), $X_{Fe}^{met}$(1) is the fraction of Fe as metal interior to distance $\delta$(1), $X_{Fe}^{met}$(2) is the fraction of Fe as metal exterior to distance $\delta$(2) and interior of distance $\delta$(3). 
The fraction of Si and Fe as metal is linearly interpolated between $\delta$(1) and $\delta$(2).
The abundance of S in each body is zero interior of distance $\delta_S$(0), equal to the CI abundance of 5.35wt\% exterior of $\delta$(3), and linearly interpolated in between.
The abundance of water in each body is zero interior of distance $\delta$(3) and 20wt\% exterior.
}
\label{tab:bestfitparameters}
\end{table}

\begin{table*}
\begin{center}
\begin{tabular}{@{}llll@{}}
\toprule
Simulation & Bulk silicate Earth & Earth-like planet & Venus-like planet \\
\midrule
\multicolumn{4}{l}{Mantle compositions} \\
SiO$_2$ & 45.40 (0.32) wt\% & 45.89 wt\% & 45.14 wt\% \\
MgO & 36.77 (0.37) wt\% & 36.82 wt\% & 35.77 wt\% \\
FeO & 8.10 (0.050) wt\% & 8.069 wt\% & 9.784 wt\% \\
Al$_2$O$_3$ & 4.49 (0.36) wt\% & 4.51 wt\% & 4.38 wt\% \\
CaO & 3.65 (0.29) wt\% & 3.62 wt\% & 3.52 wt\% \\
Na & 2590 (130) ppm & 2590 ppm & 2390 ppm \\
Cr & 2520 (250) ppm & 2800 ppm & 2990 ppm \\
Ni & 1860 (93) ppm & 1857 ppm & 3966 ppm \\
H$_2$O & 1000 (300) ppm & 1000 ppm & 60 ppm \\
S & 200 (80) ppm & 208 ppm & 1505 ppm \\
Co & 102.0 (5.1) ppm & 110.7 ppm & 197.8 ppm \\
V & 86.0 (4.3) ppm & 86.5 ppm & 91.2 ppm \\
Nb & 600 (120) ppb &  550 ppb & 590 ppb \\
Ta & 43.0 (2.2) ppb & 40.7 ppb & 39.8 ppb \\
Pt & 7.6 (1.5) ppb & 5.0 ppb & 155.1 ppb \\
Pd & 7.1 (1.4) ppb & 9.4 ppb & 108.6 ppb \\
Ru & 7.4 (1.5) ppb & 7.6 ppb & 117.3 ppb \\
Ir & 3.50 (0.35) ppb & 3.30 ppb & 79.53 ppb \\
\midrule
\multicolumn{4}{l}{Core compositions} \\
Fe & & 81.2 wt\% & 81.9 wt\% \\
Si & & 7.54 wt\% & 7.30 wt\% \\
Ni & & 5.14 wt\% & 4.90 wt\% \\
O & & 3.23 wt\% & 3.48 wt\% \\
S & & 1.89 wt\% & 1.47 wt\% \\
Cr & & 0.722 wt\% & 0.697 wt\% \\
Co & & 0.237 wt\% & 0.227 wt\% \\
V & & 119 ppm & 108 ppm \\
H & & 112 ppm & 148 ppm \\
Pt & & 5.73 ppm & 5.67 ppm \\
Ru & & 4.27 ppm & 4.23 ppm \\
Ir & & 2.90 ppm & 2.88 ppm \\
Pd & & 2.83 ppm & 2.74 ppm \\
Nb & & 551 ppb & 459 ppb \\
Ta & & 4.14 ppb & 3.73 ppb \\
\midrule
Core mass fraction & 0.32 & 0.314 & 0.300 \\
\bottomrule
\end{tabular}
\caption{The composition of the mantles and cores of the Earth-like and Venus-like planets compared to Earth (see Table 5 of \citet{Rubie:2015fj} for comparisons with other simulations). The bulk silicate Earth composition is from \citet{Palme:2003dp}.}
\label{tab:composition}
\end{center}
\end{table*}

\bibliographystyle{model2-names.bst} 
\bibliography{/Users/seth/Papers/biblio}





\end{document}